\documentstyle[11pt,newasp,twoside]{article}
\begin{document}
\title{Local Distance Indicators}
\author{Michael Feast}
\affil{Astronomy Department, University of Cape Town, Rondebosch, 7701,
South Africa. email: mwf@artemisia.ast.uct.ac.za}
\begin{abstract}
A summary is given of the calibration of the Cepheid
scale required in the derivation of distances from
observations at $V$ and $I$, as in the HST work on extragalactic
Cepheids. There is some evidence that a small metallicity correction
may be necessary in deriving distances in this way. The evidence for this
comes both from observations of Cepheids of different metallicities
and from a comparison of the LMC distance modulus derived from
Cepheids and that derived by other methods.

\end{abstract}

\section{Introduction}
   At the present time the value adopted for $H_{0}$ depends
directly on the calibration of local distance indicators. These
local calibrators should have a number of rather obvious
characteristics.\\
1. Their absolute magnitudes should be derivable with good accuracy.
The methods used to obtain these absolute magnitudes should be
based on sound empirical results and be as free as possible from
assumptions based on theory.\\
2. The absolute magnitudes of the calibrators should ideally be
free from a dependence on either the age or the metallicity of
the objects involved. But if there are age and metallicity effects,
these must be known empirically and measurable in both the
calibrating and programme stars. \\
3. The interstellar reddenings and absorptions of both calibrating
and programme stars must be measurable with good accuracy.\\
Considerable emphasis has been put on point 1 in much of the
current literature. However, points 2 and 3 are of equal importance,
especially if one is hoping to establish scales to 5 or 10 percent.
In fact, problems relating to these two latter points seem to
account for some of the conflicting scales that have been recently
proposed.\\

\section{The Cepheid Scale}
In the following it will be assumed that the primary interest at 
this symposium is the calibration of the scale to be used in
the study of Cepheids; particularly the scale
to adopt with the HST work on extragalactic Cepheids. The way in
which the HST data are analysed differs slightly from one group
of workers to another. However, the analysis is basically
equivalent to using a period-luminosity relation at
(Johnson) $V$, PL($V$), with a measured $V$ to obtain an apparent
distance modulus. The reddening, and hence the absorption, $A_{V}$, 
is derived from the $(V-I)$ colour together with a
period-$(V-I)_{0}$ relation, PC($V-I$).\\
A galactic PC($V-I$) relation can be obtained from local
Cepheids with individual reddenings derived from three colour
$BVI$ photometry, e.g.
\begin{equation}
<V>_{0} - <I>_{0} = 0.297\, \log \, P + 0.427. 
\end{equation}
This is derived from Caldwell \& Coulson (1987) (see Feast 1999, 
appendix D, where
a somewhat more accurate procedure is given). The zero-point of the
$BVI$ reddening system is set by the reddenings of Cepheids in open
clusters whose reddenings are derived from non-Cepheid cluster
members. However, it is important to note that in 
the most secure estimates of the Cepheid zero-point, 
the reddening
zero-point is immaterial so long as one is consistently using
the same zero-point for both calibrating and programme Cepheids.
This is a major advantage of Cepheids over other distance indicators
where this differential method cannot, or as yet has not been, applied. 
\\
The Cepheid PL($V$) relation can be written: 
\begin{equation}
M_{V} = -2.81\, \log \, P + \rho_{1}. 
\end{equation}
The slope of the relation is taken from the Large Magellanic Cloud
(LMC) (Caldwell \& Laney 1991). This is the only use made of
LMC data in deriving a Cepheid calibration. The slope may be
estimated in a number of ways, none of which give significantly
different values (see Feast 1999 for a discussion of this point
and for a more detailed discussion of many of the points mentioned
in the present paper). Values of $\rho_{1}$ can be obtained in a
number of ways.\\
1. The most direct, empirical method is to use parallaxes of
galactic Cepheids. A bias free analysis of Cepheid parallaxes
from the Hipparcos catalogue (ESA 1997) gives,\\
$\rho_{1}= -1.43 \pm 0.12$\\
(Feast \& Catchpole 1997, Feast 1999). This result and its bias
free nature has been confirmed either directly or through 
Monte Carlo simulations by several groups 
(Pont 1999, Lanoix et al. 1999, Groenewegen \& Oudmaijer 2000).\\
2. Proper motions from Hipparcos can be combined with radial
velocities in a statistical parallax type solution for
$\rho_{1}$. This requires a galactic model. 
Both the proper motions
(Feast \& Whitelock 1997)
and the radial velocities 
(e.g. Pont et al. 1994)
show clearly and independently the
dominant effect of differential galactic rotation on Cepheid
motions and this is then the required model. In this way 
Feast, Pont \& Whitelock (1998) obtained,\\
$\rho_{1} = -1.47 \pm 0.13$.\\
3. An estimate of $\rho_{1}$ can be obtained from pulsation
parallaxes (The Baade-Wesselink method). In most current forms
of this method a radius derived from radial velocities and
photometry is combined with a colour-surface brightness relation
to give an absolute magnitude. Laney (1998) recently obtained
results 
which imply:\\
$\rho_{1} = -1.32 \pm 0.04 \rm (internal)$\\
(see Feast 1999). 
If this method is applied consistently the internal errors
can be very small.
However, it is not yet possible to estimate
realistically possible systematic errors in the derived radii
or in the surface brightness estimates. Progress 
should be possible in this area when accurate interferometric
observations of Cepheid radii (and their variation with phase)
are obtained. 
Though even then systematic effects in the interpretation of
the results may be difficult to estimate.
Unlike the first two methods discussed above, the
reddening zero-point is of importance when absolute magnitudes 
are derived in this way.\\
4. Cepheid luminosities can be calibrated somewhat less directly
using those Cepheids that are members of open clusters.
However, the direct determination 
of distances to some open clusters by Hipparcos
has raised a number of problems (e.g. the large change in the 
distance of the Pleiades from the value inferred pre-Hipparcos).
There are indications (see, e.g. Feast 1999, van Leeuwen 2000,
Robichon et al. 2000) that these problems arise though
a combination of photometric errors, errors
in adopted reddenings and errors in assumed 
metallicity, all of which can have a significant effect because
of the steepness of the main sequence. There is also
a suggestion that the shape of the upper main sequence
as a function of age may not agree entirely with theoretical
predictions (van Leeuwen 2000).
Evidently these questions will need to be sorted out, both
for the nearby clusters with Hipparcos parallaxes and the
clusters containing Cepheids before this method of obtaining
Cepheid luminosities can be fully trusted. 
It seems best at the present time to base a cluster distance scale
on the Hyades for which there is an excellent 
Hipparcos parallax (yielding
$\rm (m-M)_{0} = 3.33 \pm 0.01$, Perryman et al. 1998) and a metallicity
([Fe/H] = +0.13) which seems well accepted (e.g. Pinsonneault et al.
1998). The metallicity corrections of these latter authors then show
that the Hyades main sequence 
in $V,(B-V)$ corresponds to that expected for
a solar metallicity cluster at $\rm (m-M)_{0} = 3.17$, or 3.12 
if the metallicity corrections of Robichon et al. (2000) are used.
A mean of 3.14 is adopted.
Since most work on Cepheids in clusters is referred to Turner's
(1979) Pleiades main-sequence we need to see how this is affected
by this Hyades result. The Pleiades-Hyades magnitude difference in 
a $V,(B-V)$ diagram, corrected for reddening but not metallicity, 
is 2.52 mag. (Pel 1985). Thus the Turner sequence is that
expected for a solar metallicity cluster at 
$\rm (m-M)_{0} = 3.14 + 2.52 = 5.66$. If we assume that the clusters
and associations containing Cepheids which are listed in Table 1
of Feast (1999) are in the mean of solar metallicity, then the 
results given there lead to:\\ 
$\rm \rho_{1} = -1.43 \pm 0.05 (internal).$\\
The external error of this result will be higher, partly due
to the uncertainties in the metallicity correction.
Note that, as with the Baade-Wesselink determinations, this result
depends on the zero-point of the reddening scale adopted, unlike
the first two results discussed in this section.\\
A straight mean of these four zero-point determinations gives:\\
$\rm \rho_{1} = -1.41$.\\ 
This value is adopted here, although it should be noted
that the two determinations which are most securely grounded
(1 and 2 above) give a slightly brighter zero-point (--1.45)
The real uncertainty in the adopted value is likely to be somewhat
less than 0.1.
The implied distance scale is about 7 percent greater than that
used by the HST Cepheid workers. But this estimate may require
revision when the details of the revised scale reported briefly
by Freedman at this meeting, are available.

\section{Tests for Metallicity Effects}
Possible metallicity effects in deriving Cepheid distances from
$V,V-I$ data were considered in some detail in Feast (1999).
There are three possible causes of such effects.\\
1. A change of temperature at a given period.\\
2. A change of atmospheric blanketing at a given temperature.\\
3. A change in bolometric luminosity at a given period.\\
Items 1 and 2 affect the bolometric corrections at the wavelengths
in question. Laney 
(1998, 1999 and private communication) discussed Baade-Wesselink
radii and colours of Cepheids in the Galaxy ($\rm [Fe/H] \sim 0$),
the LMC ($\rm [Fe/H] \sim -0.3$), and the SMC ($\rm [Fe/H] \sim -0.6$),
which lead (Feast 1999) to an effect in the derived distance moduli
(in the $V, V-I$ system) of 
$\rm \sim 0.09 \pm \sim 0.04\, mag \, [Fe/H]^{-1}$.
Kennicutt et al. (1998) found the effect to be,
$\rm 0.24 \pm 0.16\, mag \, [Fe/H]^{-1}$ from a study of Cepheids in
regions of different metallicity in M101. Both these estimates
are in the sense that without the correction the distance of a
metal-poor Cepheid would be overestimated.\\ 
Further tests can be made
by comparing the Cepheid distance modulus of the LMC, where the Cepheids
have $\rm [Fe/H] \sim -0.3$
(Luck et al. 1998), with independent estimates of the LMC
modulus. In carrying out such a test one must bear in mind that
none of the non-Cepheid distance indicators are free from problems
of one kind or another. In addition the relative reddenings of these
indicators and the Cepheids is a source of added uncertainty.
\begin{table}
\centering
\caption{Non-Cepheid LMC Moduli}
\begin{tabular}{ll}
\multicolumn{1}{c}{Method} & {Modulus}\\
&\\
\bf RR Lyraes & \\
Parallaxes & $18.70 \pm 0.22$\\
Via HB parallaxes & $18.50 \pm 0.12$\\
Via globular clusters & $18.64 \pm 0.12$\\
Via $\delta$ Sct stars & $18.62 \pm 0.10$\\
Statistical parallaxes & $18.32 \pm 0.13$\\
&\\
\bf Miras&\\
Parallaxes & $18.64 \pm 0.14$\\
Via 47 Tuc & $18.60 \pm (0.09)$\\
&\\
SN1987A ring & $18.58 \pm 0.05$\\
&\\
LMC globular clusters & $18.52 \pm 0.11$\\
&\\
Red giant clump & $(18.55) \pm (0.05$ int.)\\
&\\
Eclipsing binary & $(18.40) \pm (0.07$ int.)\\ 
\end{tabular}
\end{table}

\subsection{The RR Lyraes}
Table 1 gives the distance modulus of the LMC as derived in different
ways from RR Lyrae variables. The basic data on the (field) RR Lyraes
in the LMC are taken from Clementini et al. (2000) including the
reddenings and the mean metallicity ($\rm [Fe/H] = -1.5$). The
various absolute magnitude calibrations are as follows:\\
1. Trigonometrical parallaxes of RR Lyrae variables in the
Hipparcos catalogue (Koen \& Laney 1998). This is the most direct
calibration but has a rather large standard error.\\
2. Hipparcos trigonometrical parallaxes of Horizontal-branch
stars (Gratton 1998). This is somewhat less straight forward
than using the parallaxes of RR Lyraes themselves (see Feast 1999).\\
3. At least three groups have discussed the calibration of RR Lyrae
absolute magnitudes based on globular clusters with distances
determined by main sequence fits to subdwarfs with Hipparcos
parallaxes. A summary and revision of this method has recently
been given by Carretta et al. (2000) and their result has been used
here.\\
4. The Hipparcos parallaxes of $\rm \delta \, Scuti$ stars can be used
to derive distances to globular clusters and hence the luminosities
of the RR Lyraes (McNamara 1997). At present this method is somewhat
uncertain since it requires an extrapolation. Once $\rm \delta \, Scuti$
stars themselves are observed in the LMC the method may prove
rather valuable.\\
5. The last value in Table 1 is that derived from statistical
parallaxes of galactic RR Lyrae variables by Gould \& Popowski (1998).
Such an analysis requires one to adopt a kinematic model. 
Gould and Popowski (along with other workers) adopt a classical model for
the galactic halo in their work. However recent studies have shown
how complex the halo actually is. 
The uncertainty in this result may thus be considerably larger than
implied by the quoted standard error.\\
Estimates of RR Lyrae absolute magnitudes can also be made from
Baade-Wesselink type analyses. There are a number of problems in
doing this and a range of absolute magnitudes have been proposed
(see Feast 1999 and references there). In view of these uncertainties
the results of this method are not included here. 

\subsection{Miras}
Mira variables show a good infrared period-luminosity relation
(Feast et al. 1989) . This can be calibrated using Miras
with Hipparcos parallaxes (Whitelock \& Feast 2000). Table 1 gives the
result using the PL relation at $\rm 2.2 \mu m$ ($K$) which should be least
affected by any metallicity differences between the LMC Miras of
a given period and those used for the calibration. There is some evidence
that at a given period the LMC Miras are metal-poor compared
with galactic ones. If so, the theory of Wood (1990) indicates that
the distance modulus shown is a lower limit
(see the discussion in Feast \& Whitelock 1999). The PL relation can
also be calibrated using the infrared ($K$) magnitudes of the three
Miras in the globular cluster 47 Tuc. The Table gives the result
obtained when the distance of 47 Tuc is taken from the discussion
of Carretta et al. (2000).
\subsection{The Ring of SN1987A}
The value quoted in Table 1 is for the distance of the LMC centroid as 
derived
from the ring round SN1987A by Panagia (1998).
This assumes the ring is circular. If the ring is elliptical the
distance modulus derived would increase to a maximum of 18.64.
\subsection{Main Sequence Fitting to LMC Globular Clusters}
Johnson et al. (1999) fit main sequences of LMC globular clusters
to that of M92 to obtain a distance modulus of the LMC. The tabulated
value is based on the distance of M92 derived by Carretta et al. (2000).
It should however be noted that these latter workers indicate
that the 
derivation of the distance to this cluster is somewhat problematic.
\subsection{The Red Giant Clump}
The use of the red giant clump as a distance indicator has been
much discussed in recent times. Using this method Udalski et al.
(2000) and Stanek et al. (2000) obtained an LMC modulus of $18.24\pm 0.08$.
On the other hand Romaniello et al. (2000) found $18.59 \pm 0.09$.
The difference between these estimates is mainly due to the adoption
of different reddening corrections and different corrections for
age and metallicity effects. An extensive study of this 
latter problem has recently
been carried out by Girardi \& Salaris (2000). They find, from models,
that there are significant effects on the clump absolute magnitude due
to both age and to metallicity. Coupling these results with population
synthesis models of the LMC 
and adopting reddenings from Romaniello et al. (2000) and Zaritsky (1999)
they derive the result given in Table 1. 
As Girardi and Salaris point out the significant age and metallicity
effects reduce considerably the usefulness of the clump as a distance
indicator. The result evidently depends strongly on both stellar evolutionary
models and population synthesis work, making it less suitable
as a primary distance indicator. For this reason the result in Table 1 is
placed in brackets.
\subsection{Eclipsing Binaries}
Deriving distances from eclipsing binaries has much in common
with the determination of pulsation parallaxes by a Baade-Wesselink
type analysis. In both cases a stellar radius is combined with
an estimate of the surface brightness 
to obtain a luminosity.
In the case of eclipsing binaries the stellar radius derived depends 
amongst other things on the adopted limb darkening. 
Guinan et al. (1998) have studied the B-type binary HV2274 in the LMC.
They combine HST spectrophotometry and optical photometry with a
Kurucz model to deduce simultaneously, the law of reddening, 
the visual absorption ($A_{V}$), the metallicity ([Fe/H]),
the surface gravity, the microturbulence, and the effective
temperature. The result thus depends heavily on the model (as well
of course on the accuracy of the spectrophotometry). They obtain
a distance modulus of 18.35 for the star and from this estimate
a distance modulus of the LMC centre of 18.30. Nelson et al. (2000) 
remeasured the optical photometry with Landolt standards and increase
this later distance modulus to 18.40, the value give in Table 1. The value 
obtained
is rather sensitive to the photometry. It can be roughly estimated
that if Nelson et al. had used Cape rather than Landolt standards
(see e.g. Menzies et al. 1991) 
they would have found an even larger
distance modulus for the LMC ($\sim 18.47$).
As with the red clump distance, the eclipsing binary distance
rests on a theoretical model and cannot, at least as yet, be
considered an empirical determination. It is thus placed in 
brackets in Table 1.
\subsection{Comparison of Cepheids and other Indicators}
\begin{table}
\centering
\caption{Comparison of LMC Moduli}
\begin{tabular}{ll}
\bf Non-Cepheid & \\
RR Lyraes & 18.54\\
Miras & 18.62 \\
SN1987A & 18.58\\
LMC Globulars & 18.52\\
Red Clump & (18.55)\\
Eclipsing Binary & (18.40)\\
&\\
\bf Cepheid &\\
$V,I$, no correction & 18.66\\
with Laney correction & 18.63\\
with Kennicutt correction & 18.59\\
\end{tabular}
\end{table}
Table 2 contains mean LMC moduli as derived from each of the
indicators in Table 1. Where several estimates for a given
indicator are listed in Table 1, a straight unweighted mean
has been taken except that the trigonometrical parallax result
for RR Lyraes has been given half weight because of its large
standard error. (Note that the statistical parallax result
for the RR Lyrae stars has been given full weight although reasons were 
given in section 3.1 for regarding it with some suspicion.)\\
The first three entries of Table 2 are probably the most reliable
and their mean is 18.58. A mean of all six entries is 18.54 which is
negligibly different. These values may be compared with the
Cepheid values which are also given in Table 2. These depend on the
the use of $V, V-I$ photometry and the PL zero-point
derived  in section 2 ($\rho_{1} = -1.41$).
No metallicity correction has been applied to the first 
entry whilst the others
have been correction according to the results of Laney or
of Kennicutt et al. as discussed in section 3.
These results suggest that the LMC test provides some additional
evidence that the use of the Cepheid $V, V-I$ method is slightly
metallicity dependent in the same sense as shown by the
results of Laney and of Kennicutt et al. However, in view of the various
uncertainties an effect in the LMC of about 0.1 mag must be considered
marginal.
\section{Conclusions}
Galactic calibration of the Cepheid PL($V$) and PC($V-I$) relations
indicate that the scale used by HST workers (at least prior to mid-2000)
needs increasing by about seven percent. Various tests suggests that
there is a small metallicity effect when using $V$ and $I$ data in the manner
adopted by the HST workers.
The size of this effect is still rather uncertain. It is possibly
$\rm \sim 0.1$ or $\rm 0.2 \, mag \, [Fe/H]^{-1}$ in the sense that
the corrected distances of metal-poor Cepheids are smaller than
the uncorrected ones.
\acknowledgments
I am indebted to various colleagues, particularly
Patricia Whitelock and Dave Laney, for preprints and information
in advance of publication.

\end{document}